\documentclass[runningheads]{llncs}
\usepackage[T1]{fontenc}
\usepackage{tabularx}
\usepackage{textcomp}
\usepackage{amsmath,amsfonts}
\usepackage{booktabs}
\usepackage{graphicx}
\usepackage{subfig}

\begin{document}
\title{Research on a Driver's Perceived Risk Prediction Model
Considering Traffic Scene Interaction}
\titlerunning{Driver's Perceived Risk Prediction Model}
%
%
\author{Chenhao Yang\inst{1}\orcidID{0009-0007-3616-4014} \and
Siwei Huang\inst{1}\orcidID{0009-0002-3456-1116} \and
Chuan Hu\inst{1}\orcidID{0000-0001-5379-1561}}
\authorrunning{C.Yang et al.}
%
\institute{ School of Mechanical Engineering, Shanghai Jiao Tong University, 800 Dongchuan Rd, Minhang District, Shanghai, 200240, China\\
\email{chuan.hu@sjtu.edu.cn }}
\maketitle              
\begin{abstract}
In the field of conditional autonomous driving technology, driver perceived risk prediction plays a crucial role in reducing traffic risks and ensuring passenger safety. This study introduces an innovative perceived risk prediction model for human-machine interaction in intelligent driving systems. The model aims to enhance prediction accuracy and, thereby, ensure passenger safety. Through a comprehensive analysis of risk impact mechanisms, we identify three key categories of factors, both subjective and objective, influencing perceived risk: driver's personal characteristics, ego-vehicle motion, and surrounding environment characteristics. We then propose a deep-learning-based risk prediction network that uses the first two categories of factors as inputs. The network captures the interactive relationships among traffic participants in dynamic driving scenarios. Additionally, we design a personalized modeling strategy that incorporates driver-specific traits to improve prediction accuracy. To ensure high-quality training data, we conducted a rigorous video rating experiment. Experimental results show that the proposed network achieves a 10.0\% performance improvement over state-of-the-art methods. These findings suggest that the proposed network has significant potential to enhance the safety of conditional autonomous driving systems.

\end{abstract}
\section{Introduction}
In the development of intelligent vehicle technology, safety is the top priority\cite{chia2022risk}\cite{wang2021towards}. According to the World Health Organization, road traffic accidents claim 1.3 million lives annually—equating to more than two deaths per minute. These accidents are largely due to unsafe driving behaviors, often arising from driver's inability to properly perceive risks in their surroundings\cite{skrickij2020autonomous}. Thus, assessing driver risk perception and developing intelligent systems that accurately perceive the environment and predict risks have become critical challenges for ensuring passenger safety\cite{10266784}\cite{10345740}. This capability, known as "Perceived Risk Prediction," has gained increasing attention in recent years\cite{kolekar2020human,song2024subjective,huang2020integrated}. Perceived risk prediction not only supports stability control systems to ensure safety but also enhances the driving experience\cite{10038483}. By aligning risk assessment with driver's expectations, it enables a more human-like autonomous driving experience, improving both safety and comfort\cite{song2023personalized}.

Developing an effective perceived risk prediction model faces two main challenges: data-related issues, such as selecting and modeling the right data (where both insufficient and excessive data can hurt accuracy), and methodological issues, including developing models that optimize prediction performance. The following sections review the current research landscape and introduce the contributions of our study.

\subsection{Current Research on Data for Perceived Risk Modeling}
A driver’s perceived risk refers to their recognition of hazardous situations in road traffic environments\cite{swain2021motion}. However, due to its complex influencing factors, quantifying it mathematically or physically is challenging. Research typically focuses on three categories: driver's personal characteristics (subjective), ego-vehicle motion (objective), and surrounding environment characteristics (objective).

Studies have shown that personal characteristics like gender, driving experience, and age significantly influence perceived risk\cite{machado2016socio}. Previous research has analyzed the impact of these factors on perceived risk, with some studies incorporating traits such as gender, age, and automation capability into risk models\cite{borowsky2010age}\cite{rhodes2011age}. This study also considers these characteristics and the challenges in data collection.

In contrast, modeling perceived risk using ego-vehicle motion and environmental factors focuses on objective elements related to collision risk. Models like PPDRF, PCAD, and PODAR use motion states (e.g., speed, direction, acceleration) and road conditions in risk assessment\cite{tan2021risk,he2024new,chen2022quantifying}. Bao et al. expanded this by adding vehicle kinematic and dynamic features (e.g., TTC, braking force, steering angle)\cite{bao2020personalized}. Ping et al. integrated environmental factors like weather and road conditions\cite{9827058}.

While these factors provide unique insights into perceived risk modeling, current research has not fully explored the complementarity of diverse data types. This study combines subjective and objective factors to analyze the risk impacts of driver characteristics, ego-vehicle motion, and environmental elements.

To measure perceived risk, methods like questionnaires\cite{wang2020driving}, indirect expressions (e.g., gaze duration, reaction time)\cite{jiang2022risk}, and driving video scoring are used\cite{9827058} Although questionnaires offer detailed insights, they are time-consuming and impractical for real-time prediction. Indirect expressions are informative but lack a direct link to actual driver experiences\cite{wetton2010development}. Driving video scoring, based on real-world observations, has proven to be a direct and effective method for assessing perceived risk\cite{bao2020personalized}\cite{9827058}.

\subsection{Current Research on Perceived Risk Modeling Methods}
Traditional approaches to perceived risk modeling primarily follow two methods. The first method constructs physical or mathematical models based on assumptions about traffic environments or participants\cite{8574961}, such as the PODAR model by Chen et al., which uses time and distance decay functions to estimate collision risks\cite{chen2022quantifying}. These models offer good interpretability due to their rigorous logical foundations but face limitations in data selection, making it difficult to fully capture the driver’s perceived risk.

The second method employs machine learning techniques, such as Bayesian inference and hidden Markov models, to establish relationships between influencing factors and perceived risk\cite{zheng2020novel}\cite{shi2019feature}. These models achieve certain prediction accuracies by learning from large datasets. However, due to their relative simplicity, they struggle to capture the complex interactions between factors, limiting prediction accuracy.

In contrast to these traditional methods, deep learning offers greater flexibility and better fitting capabilities. LSTM and GRU networks, designed for sequential data with memory mechanisms, are particularly suitable for time-series tasks\cite{gandrez2023predict,bao2019personalized,9827058}. While deep learning methods demonstrate superior performance, existing models often neglect the interactions among traffic participants and suffer from poor interpretability. To address these issues, this study proposes a perceived risk prediction model that integrates deep learning with risk field theory and incorporates interaction relationships among participants using a cross-attention mechanism, enhancing both prediction accuracy and interpretability.
\subsection{Contributes of This Study}
In summary, this study addresses the limitations identified in existing research and proposes an enhanced model for perceived risk prediction, incorporating interactions within traffic scenarios. The main contributions of this work are as follows:

1. Data Integration: Subjective and objective factors, including ego-vehicle motion, surrouning environment characteristics and driver's personal characteristics, were comprehensively considered. Risk impact mechanisms and variance analyses were performed to identify optimal modeling features, thereby improving predictive performance.

2. Prediction Methods: A framework combining risk field theory and deep learning was developed. A neural network incorporating LSTM and cross-attention mechanisms was designed to model traffic interactions. Additionally, a personalized modeling strategy was implemented to further enhance prediction accuracy.

3. Experimental Design:  We conducted a scenario-rich driving video scoring experiment with sufficient data, ensuring the model’s high accuracy and broad applicability. 

The remainder of this paper is organized as follows: Section 2 presents the framework and details of the perceived risk prediction model proposed in this study. Section 3 describes the experimental design. Section 4 provides an analysis of the experimental results. Section 5 concludes the study.

\section{The Structure of The Perceived Risk Prediction Model}
As shown in Figure 1, the proposed model consists of three key components. First, features are selected from ego-vehicle motion, surrounding environment, and driver's personal characteristics through risk impact mechanism analysis. Next, a clustering-based personalized modeling method classifies drivers. Finally, the driving data from each driver group trains a deep learning network combined with Long Short-Term Memory (LSTM) networks, a Cross-Attention mechanism, and a risk field model, resulting in the final perceived risk prediction model. The following sections will detail the feature selection process, the personalized modeling strategy and the network architecture of the perceived risk prediction model.

\begin{figure}[!t]
\centering
\includegraphics[width=\textwidth]{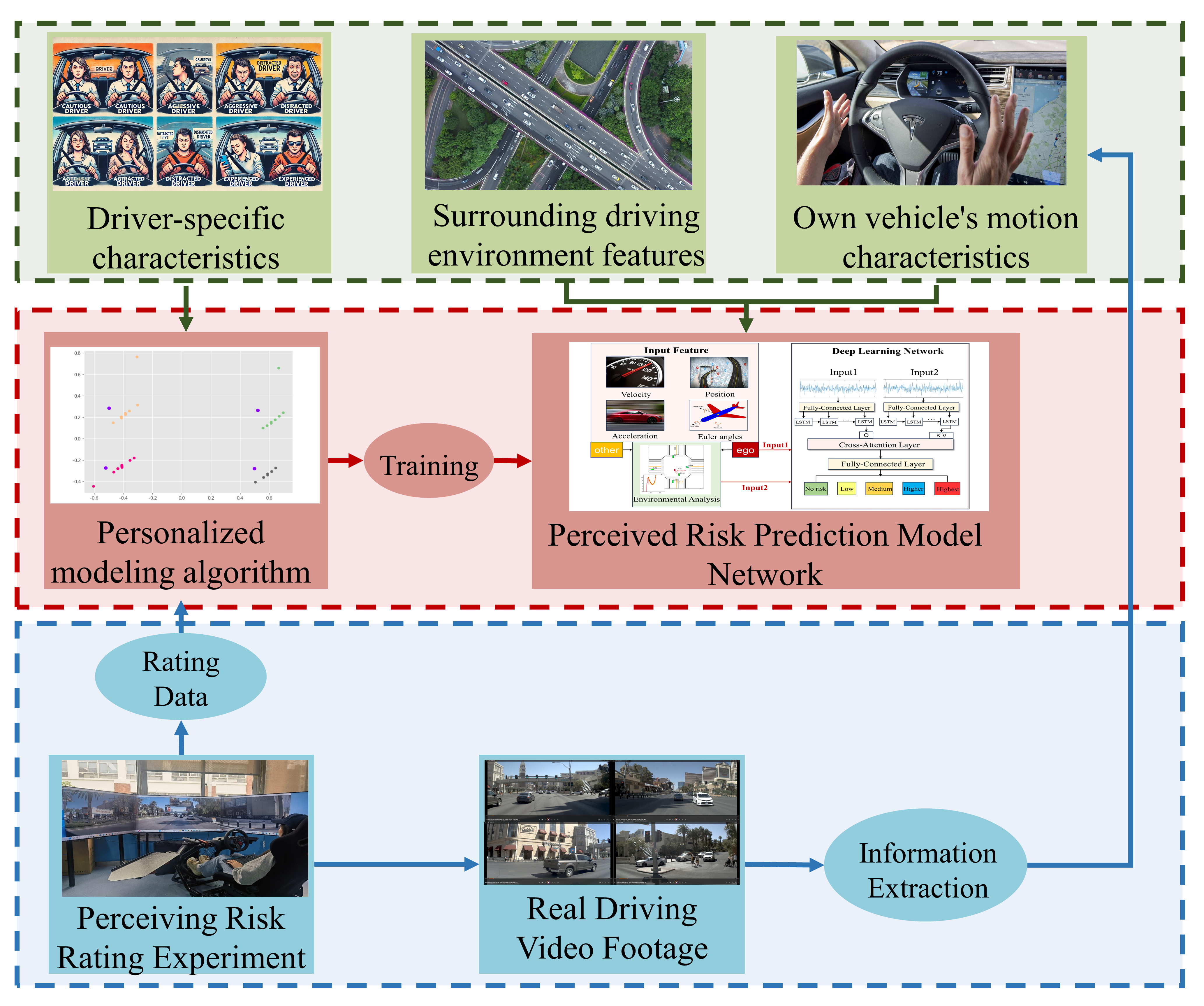}
\caption{Perceived risk prediction model framework.} \label{fig1}
\end{figure}
\subsection{Feature Selection}
Perceived risk refers to a driver's judgment of potential risks in the road traffic environment, influenced by both objective risks and subjective perception. Based on the analysis of potential risk\cite{he2024new}, this study proposes two hypotheses:

1. Drivers assess the difficulty of avoiding collisions based on their visual perception of the relative motion between their vehicle and adjacent vehicles, which affects their risk perception.

2. Motion uncertainties of both the subject vehicle and surrounding vehicles contribute to additional perceived risk, as drivers predict trajectories to estimate collision risks.

Thus, potential risk is closely linked to collision risk and the motion state information of traffic participants. Objective modeling features include the two-dimensional speed, position, acceleration, and Euler angles of the ego-vehicle, as well as similar dynamic features of surrounding vehicles. The number and types of surrounding traffic participants are also considered.

Driver's personal characteristics further influence perceived risk. Researches have shown that factors such as age, gender, and driving experience significantly impact perceived risk \cite{wallis2007using}\cite{harre2005self}. Based on these findings, the study incorporates driver characteristics (gender, age, driving experience, and style) into the personalized modeling process.

All modeling features are summarized in Table 1.
\begin{table}
  \centering
  \caption{Summary of modeling features.}\label{table1}
  \renewcommand\arraystretch{1.3}
  \begin{tabular}{ m{5cm}<{\centering} m{7cm}<{\centering} }
    \hline
       \textbf{Data Categories} & \textbf{Specific Features}  \\
       \hline
       Ego-vehicle motion features & Velocity, Acceleration, Position, Euler angles  \\
       \hline
       Surrounding environment characteristics & Same as above, Categories and number of traffic participants  \\
       \hline
       Driver's personal characteristics & Gender, Age, Driving experience, Driving style \\
       \hline
  \end{tabular}
\end{table}
\subsection{Personalized Modeling Strategy}
Driver's personal characteristics significantly influence their risk perception abilities. Therefore, incorporating personalized features based on these characteristics can improve the accuracy of risk perception predictions and enhance safety.

We propose a personalized modeling approach using clustering to classify drivers based on their individual characteristics. By grouping drivers with similar traits, we ensure that those with comparable risk perceptions are classified together. This method improves predictive accuracy over aggregated driver data and enhances generalizability compared to models focused on individual drivers.

The classification considers four dimensions: Gender, Age, Driving Experience, and Driving Style. Gender and driving style are encoded numerically (male = 1, female = 0; aggressive = 2, moderate = 1, conservative = 0). The feature vector thus has four dimensions. The clustering process involves the following steps:

1. Preprocessing Data: Normalize the input feature vectors $\{Y_1, Y_2, \ldots, Y_n\}$ after removing outliers. Each $Y_i$ is a $1 \times 4$ column vector $\{y_{i1}, y_{i2}, y_{i3}, y_{i4}\}^T$. Normalization is applied as:
\begin{equation}
\label{eq4}
y_{ij} = \frac{y_{ij}}{\max_{l,h} |y_{lj} - y_{hj}|}
\end{equation}

2. Clustering Initialization: Define $p$ clusters (starting from $p=1$), randomly selecting $p$ points as initial cluster centers $\{\mu_1, \mu_2, \ldots, \mu_p\}$.

3. Loss Function Optimization: Minimize the distance between sample points and their corresponding cluster centers using the loss function:
\begin{equation}
\label{eq5}
J(c, \mu) = \min \sum_{i=1}^n d_2(Y_i, \mu_{c_i})
\end{equation}
\begin{equation}
\label{eq6}
d_2(Y_i, Y_j) = \sum_{k=1}^4 \sqrt{y_{ik}^2 - y_{jk}^2}
\end{equation}

4. Iterative Training: Assign points to clusters based on minimal distance:
\begin{equation}
\label{eq7}
c_i^t \leftarrow \arg\min_k d_2(Y_i, \mu_k^t)
\end{equation}
Update cluster centers:
\begin{equation}
\label{eq7}
\mu_i^{t+1} \leftarrow \arg\min \sum_{i=1}^n d_2(Y_i, \mu_{c_i})
\end{equation}

5. Determine Optimal Clusters: Increment $p$ and repeat clustering. Evaluate results using metrics like the silhouette coefficient, average deviation, and SSE (Sum of Squared Errors) to select the best number of clusters.
\subsection{Network Architecture of the Perceived Risk Prediction Model}
We propose a perceived risk prediction model based on a deep learning network, as shown in Figure 2. The objective features are divided into two categories: ego-vehicle motion(Category A) and surrounding environmental characteristics(Category B), sampled at a frequency of 10Hz. To ensure the interpretability of the model, we designed an environmental analysis algorithm based on Categories A and B to generate environmental risk features (Category C). Additionally, we developed a deep learning network capable of considering the interaction between the ego-vehicle and the external environment, using Categories A (Input 1) and C (Input 2) as inputs to predict the perceived risk score. To ensure prediction accuracy, this study defines five risk levels: no risk (0), low risk (1), medium risk (2), higher risk (3), and highest risk (4).
\begin{figure}
\centering
\includegraphics[width=\textwidth]{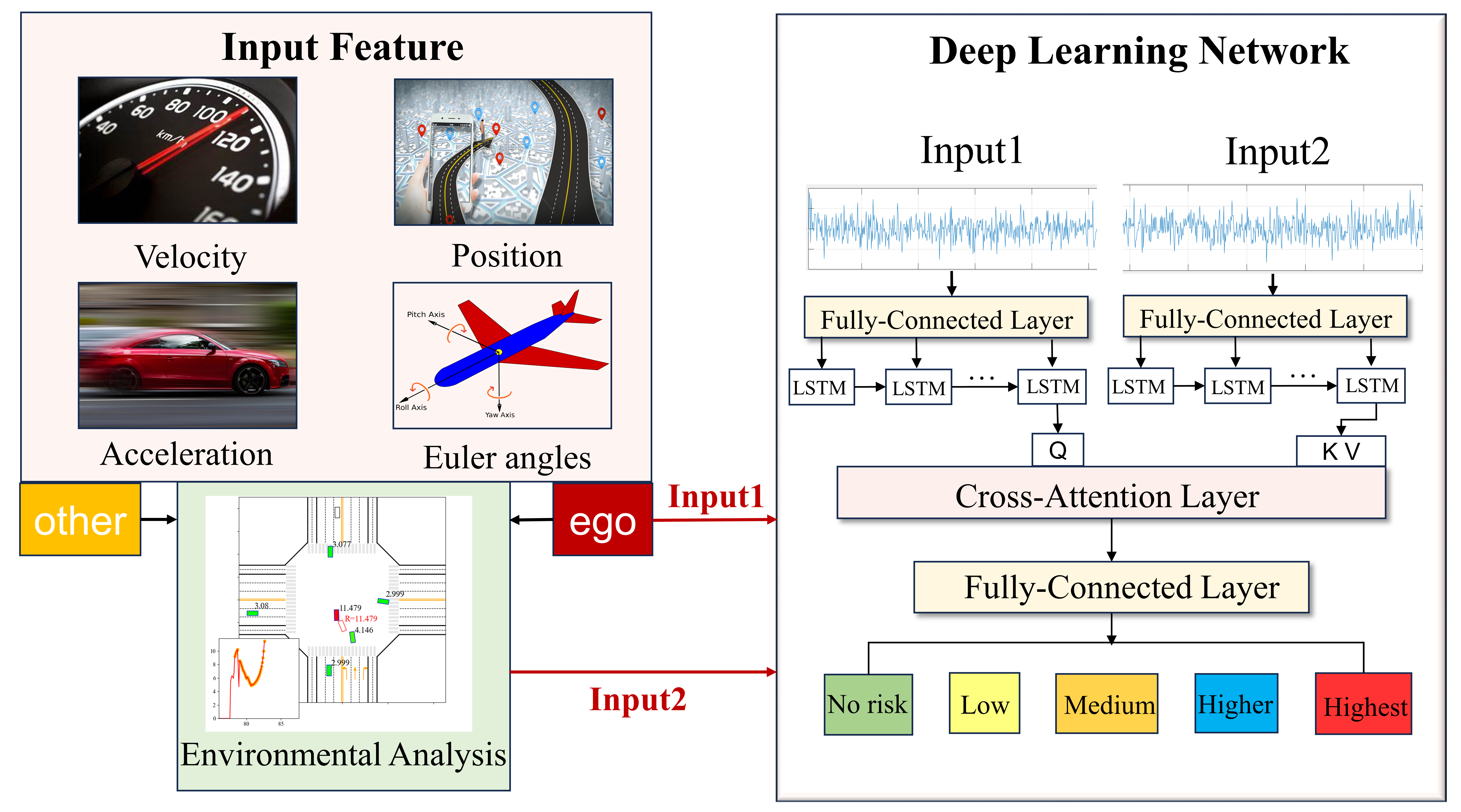}
\caption{Perceived risk prediction network structure.} \label{fig1}
\end{figure}

We use the PODAR risk field model to design the environmental analysis algorithm, ensuring the stability of feature dimensions, simplifying the processing, and enhancing the interpretability of environmental risk features\cite{chen2022quantifying}. The process of the environmental analysis algorithm is as follows:

1. Risk Value Calculation: For each traffic participant within the detectable range, calculate the PODAR risk value, which incorporates potential collision risk and considers time- and distance-based decay factors. The key inputs for calculation include absolute and relative speeds, relative positions, and participant types. The calculation formula is as follows:
\begin{equation}
\label{eq1}
G_t = \frac{1}{2} \cdot (M^{\text{ego}} + M) \cdot V_t \cdot \lvert V_t \rvert
\end{equation}
\begin{equation}
\label{eq2}
\text{PODAR}_t^n = G_t^n \cdot \omega_D \cdot \omega_T
\end{equation}
Where $G_t$ is the potential collision function, $M$ represents the virtual mass, and $V_t$ indicates the relative motion between the ego vehicle and other traffic participants.

2. Objective Risk Division: Divide the overall risk into four regions based on the driver's front, rear, left, and right viewpoints. For each region, the maximum PODAR risk value is calculated:
\begin{equation}
\label{eq3}
\text{Risk}_i = \max \{ \text{PODAR}_t^n(i) \}
\end{equation}
Where $i$ represents the viewpoints (front, rear, left, right), and $\text{Risk}_i$ is the maximum risk value for that region.

3. Weighted Risk Calculation: To include critical features, such as the number of traffic participants, a weighted calculation method is used. Traffic participants are categorized as vehicles and pedestrians, with viewpoint-based weights assigned (front = 1, left/right = 0.6, rear = 0.3). These weights are applied to the participant counts, resulting in a weighted sum. The extracted risk features include the PODAR risk values for four viewpoints and the weighted counts of vehicles and pedestrians.

These features, along with the ego-vehicle motion, are input into the deep learning network for perceived risk prediction. Unlike previous methods that concatenated ego-vehicle and environmental data into a single vector, this model uses LSTM to handle sequential data and incorporates a cross-attention mechanism to capture the interaction between the ego-vehicle and the environment. The processed features are input into the “query” and “key/value” networks of the cross-attention mechanism, further enhancing the model's ability to predict perceived risk across five risk levels.
\section{Experiment Design}
We designed a video scoring experiment. Volunteers watched real driving video footage and provided frame-by-frame perceived risk ratings using a custom Python script. This method collected approximately 14,600 data points from 30 minutes of footage, significantly exceeding traditional questionnaire methods and meeting the data volume requirements for deep learning training. To simulate real driving, screens were arranged to mimic in-car views, and volunteers were trained to adopt the driver’s perspective. The volunteers can experience a variety of driving scenarios during the experiment, ensuring the model's broad applicability.
\subsection{Objective and Deployment}
The experiment utilizes the nuPlan dataset, displayed on a three-screen setup in the lab. The dataset offers detailed semantic annotations, distinguishing vehicles, pedestrians, and obstacles. It includes motion state data for the ego vehicle and other traffic participants, such as speed, acceleration, position, and Euler angles, which meet the study's modeling requirements. The video footage covers front, left, right, and rear views, providing volunteers with an immersive driving experience.

To simulate realistic driving scenarios, we use a triple-screen setup displaying videos from four viewpoints: front, left, right, and rear. As shown in Figure 3, the front-view video is on the center screen, while the rear-view video, displayed on a smaller interface, is positioned at the top center of the front screen to simulate a rearview mirror. The left and right side-view videos are shown on the left and right screens, slightly overlapping with the edges of the front view to create a cohesive visual experience.
\begin{figure}[!t]
\centering
\includegraphics[width=0.9\textwidth]{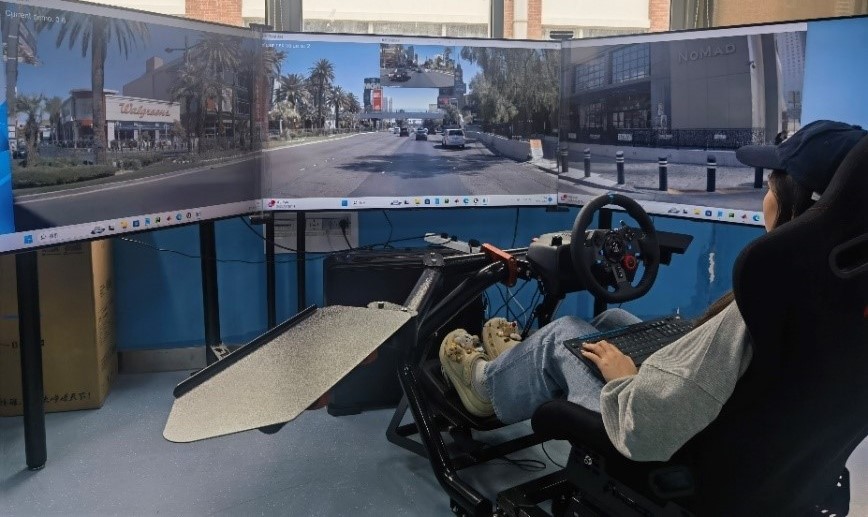}
\caption{Experiment process display.} \label{fig1}
\end{figure}
\begin{figure}[!t]
\centering
\includegraphics[width=0.9\textwidth]{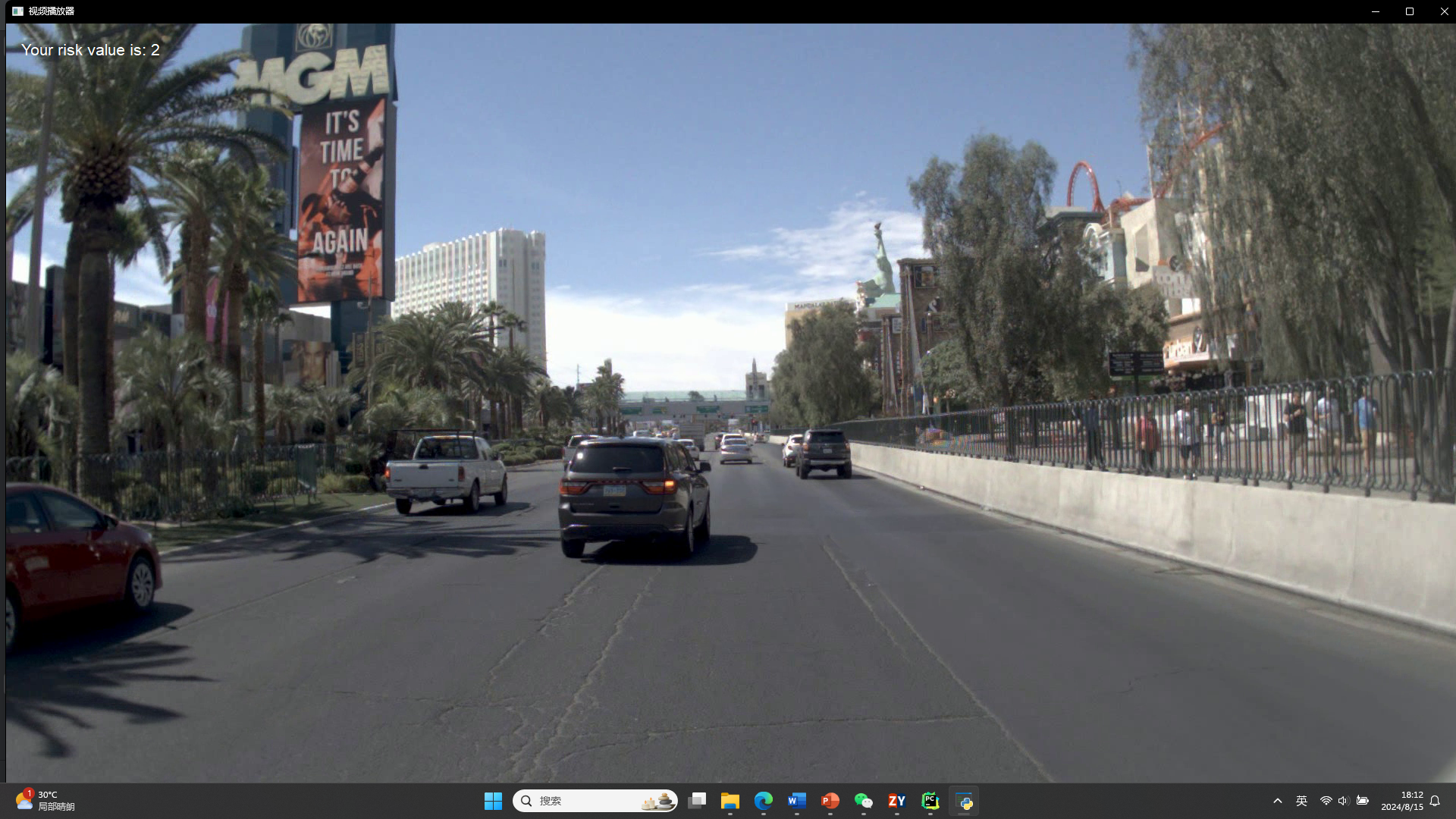}
\caption{Rating and video material interface.} \label{fig1}
\end{figure}
\subsection{Detailed Design}

To ensure sample diversity while maintaining volunteer rating effectiveness, a preliminary demo selection process was conducted. The total duration of the demo was limited to 30 minutes, as volunteers can maintain focus for this period. The selected demo lasts 1,460 seconds (about 28 minutes, including interaction time) and features diverse traffic scenarios, such as straight driving, side overtaking, deceleration near traffic lights, waiting at lights, turning at intersections, and slowing or stopping in specific road segments.

The demo is displayed on a triple-screen setup, with viewpoints adjusted to simulate real driving conditions. During the experiment, participants use keyboard keys (0-4) to rate risk levels.

The experiment involved 42 participants (26 males, 16 females) with an average driving experience of 2.125 years (range: 1 to 6 years). Informed consent was obtained from all participants. Before the main experiment, participants watched two test videos to familiarize themselves with the rating process and set standards. Researchers reminded them to focus on risk factors such as vehicle distance and speed from the driver’s perspective. As shown in Figure 4, the rating interface and video (front view only) are displayed together, with the result shown in the top left corner to avoid obstructing the view.

\section{Analysis and Discussion of Experimental Results}
This section presents the experimental results, including network training and result validation. First, a significance analysis of the impact of modeling features on perceived risk was conducted. Then, the effectiveness of the personalized modeling strategy and the LSTM network architecture combined with the cross-attention mechanism was validated.
\subsection{Significance Analysis of The Impact of Modeling Features on Perceived Risk}
This study used analysis of variance (ANOVA) to evaluate the significance of various selected features on driver's risk perception and adjusts the feature selection process accordingly. The results are presented in Table 2 and Figure 5, respectively.
\begin{table}
  \centering
  \caption{Anova Results for Different Features}\label{table4}
  \renewcommand\arraystretch{1.2}
  \begin{tabular}{ccccc}
 \toprule
    \textbf{Features} & \textbf{Sumsq}  & \textbf{F} & \textbf{P}\\
    \midrule
       $Velocity$ & 0.467 & 2.883 & 0.0920  \\
       $Acceleration$ & 1.354 & 8.353 & 0.0045  \\
       $Risk_{front}$ & 2,051 & 12.655 & 0.0010  \\
       $Risk_{left}$ & 0.982 & 6.604 & 0.0152 \\    
       $Risk_{right}$ & 1.077 & 6,646 & 0.0111 \\ 
       $Risk_{back}$ & 0.461 & 2.845 & 0.0941 \\
       $Number_{pe}$ & 0.387 & 2.389 & 0.1248 \\
       $Number_{ve}$ & 0.471 & 2.908 & 0.0907 \\
    \bottomrule
  \end{tabular}
\end{table}
\begin{figure}[!t]
\centering
\includegraphics[width=0.9\textwidth]{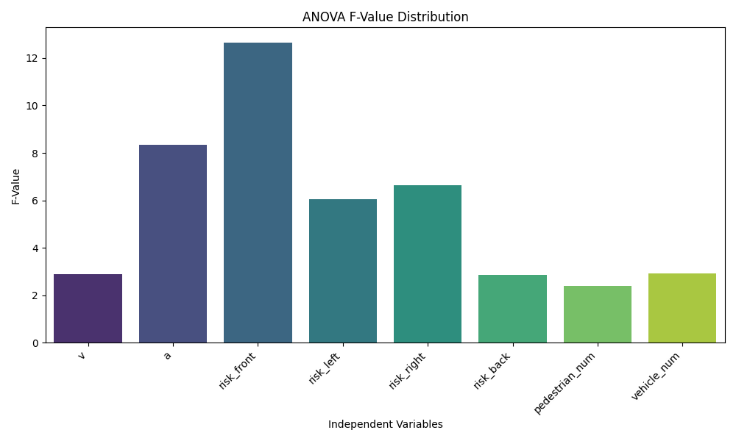}
\caption{Rating and video material interface.} \label{fig1}
\end{figure}
As shown in Table 2, eight features were analyzed: velocity, acceleration, front risk, left risk, right risk, rear risk, pedestrian count, and vehicle count. Based on the ANOVA results, the p-values for different features were below or close to 0.1, indicating their significant impact on perceived risk. Acceleration and the left, right, and front risk features had p-values below 0.05, with F-values exceeding 6, demonstrating a highly significant influence. These risk features, derived from the environmental analysis algorithm, validate its effectiveness in providing interpretable features and optimizing the predictive performance of the deep learning model.

The study found that drivers are more sensitive to acceleration than speed, with acceleration being more likely to trigger higher perceived risk. Among directional risks, drivers focused most on the front environment, followed by the left and right sides, and least on the rear. Regarding pedestrian and vehicle counts, drivers were more sensitive to the vehicle count, likely because the presence of pedestrians is critical, but their exact number is less relevant.

These findings provide guidance for feature selection and parameter settings in deep learning networks, as well as valuable insights for intelligent driving systems. For example, reducing acceleration, rather than simply lowering speed, may more effectively reduce perceived risk while improving driving efficiency.
\subsection{Validation of the Effectiveness of the Deep Learning Network Architecture}
This section evaluates the model's effectiveness and generalization ability by calculating and analyzing the AUC (Area Under the Curve) values of various classification networks. AUC provides a comprehensive performance assessment across different classification thresholds and is unaffected by class imbalance.

The deep learning network proposed in this study combines LSTM with a cross-attention mechanism (LSTMCA), effectively capturing interactions between traffic participants to improve prediction performance. According to recent research by Gandrez et al. (2023), LSTM-based models perform well in predicting perceived risk. This section compares the LSTMCA model with a traditional LSTM model and five classic machine learning algorithms: Support Vector Machine (SVM), Regularized Support Vector Machine (RBF-SVM), Linear Discriminant Analysis (LDA), Quadratic Discriminant Analysis (QDA), and Fully Connected Neural Network (FCNN).

In model construction, the classification network in the algorithm was replaced with the above six network structures. Seven groups of network models were obtained: LSTMCA, LSTM, SVM, RBF-SVM, LDA, QDA, and FCNN. All models were trained using the complete scoring dataset from all volunteers. It is important to note that traditional models like SVM, LDA, and QDA are better suited for fitting simpler functional mappings, while FCNN and LSTM models do not fully capture the interactions in traffic scenarios, which makes them less adaptable to complex driving contexts compared to the model proposed in this study.

 Table 3 lists the AUC values for all model groups. Figure 6 presents the classification performance of the LSTM and LSTMCA models on the test set using confusion matrices, with the vertical axis representing the true risk scores and the horizontal axis representing the predicted results.
 \begin{table}
  \centering
  \caption{AUC Results for Different Model Groups}\label{table5}
  \renewcommand\arraystretch{1.2}
  \begin{tabular}{cccccccc}
 \toprule
    \textbf{SVM} & \textbf{RBFSVM} & \textbf{LDA} & \textbf{QDA} & \textbf{FCNN} & \textbf{LSTM} & \textbf{LSTMCA}\\
    \midrule
       0.632 & 0.643 & 0.591 & 0.604 & 0.753 & 0.809 & \textbf{0.895} \\
    \bottomrule
  \end{tabular}
\end{table}
\begin{figure}
\centering
\subfloat[]{\includegraphics[width=0.45\textwidth]{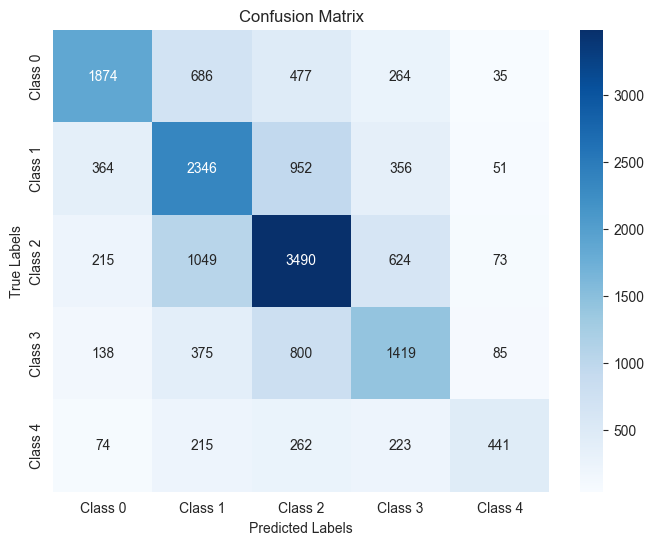}%
\label{fig81}}
\hfil
\subfloat[]{\includegraphics[width=0.45\textwidth]{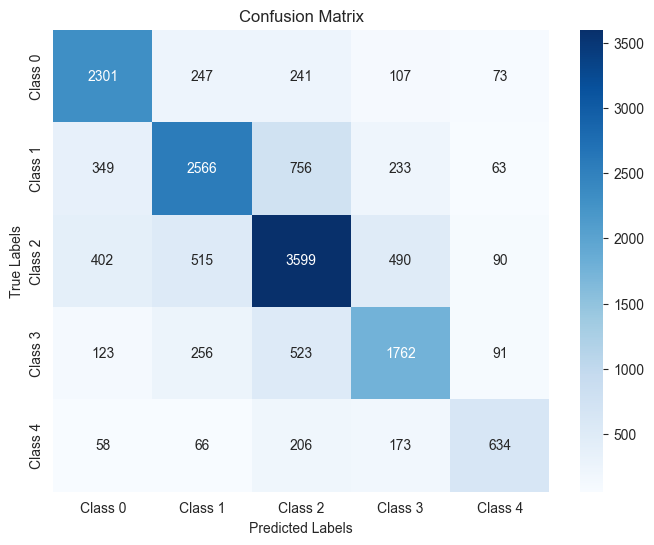}%
\label{fig82}}
\caption{Comparison of confusion matrices between the LSTM group and the LSTMCA group.(a) LSTM, (b) LSTMCA }
\label{fig8}
\end{figure}
From Table 3, the AUC values for the LSTM and LSTMCA groups are significantly higher than those of the other five groups, highlighting the superior performance of LSTM-based networks in risk prediction. The results further show that the LSTMCA group outperforms the LSTM group. Specifically, the LSTM group achieves an AUC of 0.809, while the LSTMCA group reaches 0.895, reflecting a 10.9\% improvement. This demonstrates the effectiveness of the proposed network architecture, particularly the cross-attention mechanism.

This section validates the effectiveness of the LSTM with Cross-Attention architecture while excluding the influence of personalized modeling strategies. Section 4.2 will further evaluate the network architecture's effectiveness under personalized modeling strategies and demonstrate the full predictive capability of the proposed risk prediction model.
\subsection{Validation of the Effectiveness of the Personalized Modeling Strategy}

In this study, 42 participants were recruited, and 40 valid datasets were collected. The personal characteristics of the drivers in these datasets were statistically analyzed and classified using the personalized modeling strategy. The optimal number of clusters was determined to be 4 based on the average deviation method and silhouette coefficient method. To visualize the results, Principal Component Analysis (PCA) was used to reduce the four-dimensional data to two dimensions, and the clustering results are shown in Figure 7.
\begin{figure}
\centering
\includegraphics[width=0.7\columnwidth]{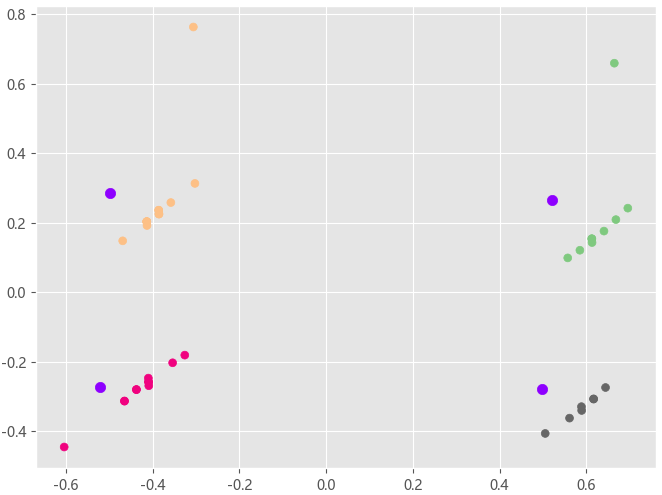}
\caption{Cluster classification result plot.}
\label{fig_10}
\end{figure}

In Figure 7, the purple dots represent the cluster centers, while the other colors indicate the four-category classification results of the 40 sample data points. After classification, the driver feature data are grouped and concatenated, then trained in their respective personalized networks.

The volunteer data were divided into four groups: Category 1, Category 2, Category 3, and Category 4. Four perceived risk prediction networks were trained for these groups, classified on the test set, and their AUC values were calculated. These results were compared to those from non-classified data to validate the effectiveness of the personalized modeling strategy. Additionally, data from different groups were trained on six control networks, further confirming the superiority of the proposed model.

Table 4 presents the AUC results, while Figure 8 illustrates the classification performance of the LSTMCA model across different groups using confusion matrices.
\begin{table}
  \caption{AUC Statistical Results for Different Networks and Groups}\label{table6}
  \renewcommand\arraystretch{1.2}
  \begin{tabular}{m{1.8cm}<{\centering} m{1.3cm}<{\centering} m{1.5cm}<{\centering} m{1.3cm}<{\centering} m{1.3cm}<{\centering} m{1.4cm}<{\centering} m{1.4cm}<{\centering} m{1.5cm}<{\centering}}
 \toprule
    \textbf{Group} & \textbf{SVM} & \textbf{RBFSVM} & \textbf{LDA} & \textbf{QDA} & \textbf{FCNN} & \textbf{LSTM} & \textbf{LSTMCA}\\
    \midrule
       All & 0.632 & 0.643 & 0.591 & 0.604 & 0.753 & 0.809 & \textbf{0.895} \\
       Category1 & 0.674 & 0.701 & 0.629 & 0.657 & 0.823 & 0.863 & \textbf{0.941}\\
 Category2 & 0.689 & 0.697 & 0.651 & 0.664 & 0.839 & 0.857 & \textbf{0.945} \\
       Category3 & 0.687 & 0.681 & 0.618 & 0.643 & 0.851 & 0.882 & \textbf{0.958} \\
       Category4 & 0.641 & 0.653 & 0.642 & 0.639 & 0.803 & 0.848 & \textbf{0.950} \\
       Average & 0.673 & 0.683 & 0.635 & 0.651 & 0.829 & 0.863 & \textbf{0.949} \\
    \bottomrule
  \end{tabular}
\end{table}
\begin{figure}[!t]
\centering
\subfloat[]{\includegraphics[width=0.45\columnwidth]{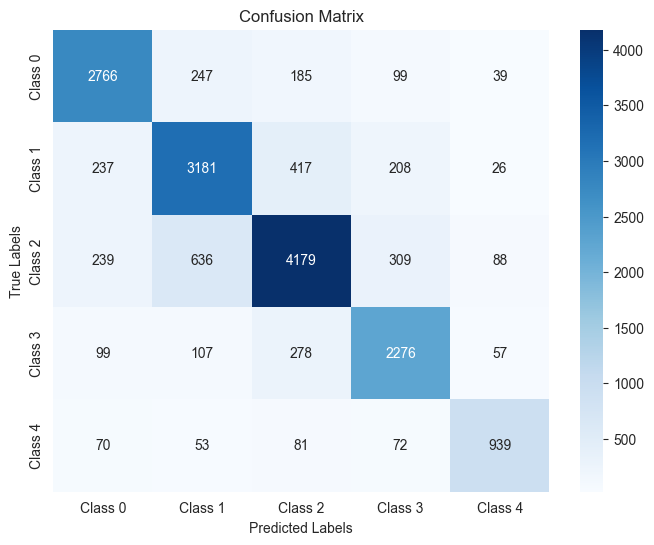}%
\label{fig3}}
\hfil
\subfloat[]{\includegraphics[width=0.45\columnwidth]{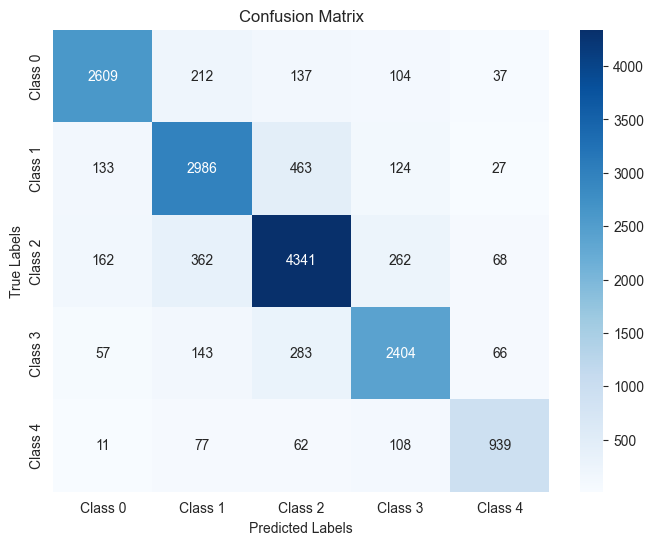}%
\label{fig4}}
\quad
\subfloat[]{\includegraphics[width=0.45\columnwidth]{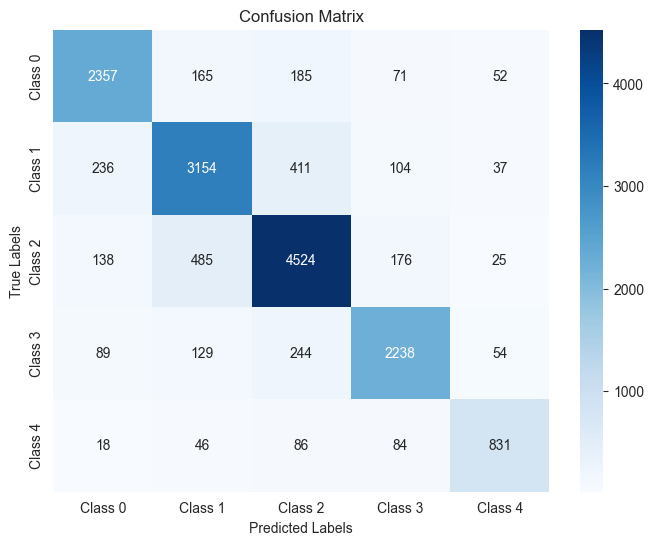}%
\label{fig5}}
\hfil
\subfloat[]{\includegraphics[width=0.45\columnwidth]{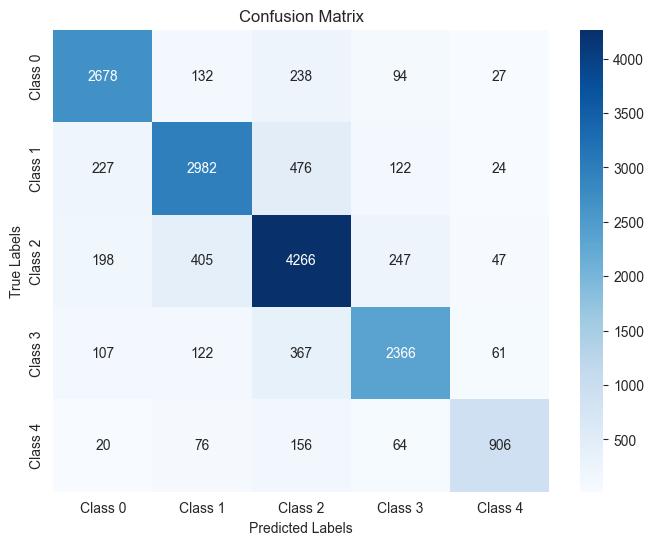}%
\label{fig6}}
\caption{Confusion matrices for LSTMCA network classification in the four groups.(a) Category1, (b) Category2, (c) Category3, (d) Category4 }
\label{fig11}
\end{figure}

Comparing the experimental results of the four classification groups in the LSTMCA model with the overall driver group, the AUC increased from 0.895 to 0.949, a notable improvement of 6.0\%. In Category 3, the AUC rose even further by 7.0\%, confirming the effectiveness of the personalized modeling strategy. The other six models showed AUC improvements ranging from 6.2\% to 9.3\%, further validating the approach.

When comparing classification performance across different network structures, the ranking is as follows: LSTMCA > LSTM > FCNN > RBF-SVM > SVM > QDA > LDA. The LSTMCA model consistently achieved high performance, with AUC values exceeding 0.91 for all sample groups. Compared to the LSTM model, the AUC increased from 0.863 to 0.949, a significant 10.0\% gain, highlighting the effectiveness of the network architecture and the superiority of the proposed risk prediction model.

Further analysis of the LSTMCA model’s classification results includes calculations of precision, recall, and F1-Score for each class. The average values are summarized in Table 5.

The average values are summarized in Table 5. Most of the precision, recall, and F1-Score results exceed 0.8, with the highest exceeding 0.83, demonstrating excellent performance in the challenging five-class classification task. Among the precision, recall, and F1-Score results across different risk levels: Risk Level 1 performed relatively poorly, while Risk Level 0 performed the best. However, the performance differences across risk levels were minimal, highlighting the model's ability to classify all levels consistently and avoid performance imbalance.
\begin{table}
  \centering
  \caption{Analysis Results of Precision, Recall, And F1-Score for Each Class}\label{table7}
  \renewcommand\arraystretch{1.2}
  \begin{tabular}{cccccc}
 \toprule
    \textbf{Measures} & \textbf{Risk0} & \textbf{Risk1} & \textbf{Risk2} & \textbf{Risk3} & \textbf{Risk4}\\
    \midrule
       Precision & 0.837 & 0.784 & 0.810 & 0.806 & 0.834  \\
        Recall & 0.837 & 0.789 & 0.818 & 0.804 & 0.770 \\
       F1-Score & 0.837 & 0.787 & 0.814 & 0.805 & 0.800 \\
    \bottomrule
  \end{tabular}
\end{table}
\section{Conclusion}

Safety is crucial in intelligent vehicles, and perceived risk prediction helps drivers identify potential hazards and take evasive actions to ensure passenger safety. This study focuses on constructing an accurate perceived risk prediction model.

Based on the risk impact mechanism, three categories of modeling features are identified: driver's personal characteristics, ego-vehicle motion, and surrounding environment characteristics. A personalized modeling method for driver's personal characteristics is designed, and a deep learning framework incorporating traffic scenario interactions is applied for risk prediction.

In the experimental section, a real driving video library was constructed using the nuPlan dataset. Training data were collected by recruiting volunteers to watch videos and rate the risks. Ablation experiments verified the accuracy of the LSTM model with a cross-attention mechanism, while AUC-based comparisons of machine learning methods and network models further validated the effectiveness of the personalized modeling strategy.

The proposed perceived risk prediction model provides key safety insights for both drivers and intelligent vehicle systems. For drivers, a human-machine interaction system can issue warnings (such as voice alerts or seat vibrations) when perceived risk exceeds a threshold, improving reaction time. For intelligent vehicles, safety control algorithms based on risk perception results can dynamically adjust speed, acceleration, and steering, ensuring safety and enhancing driving comfort \cite{10164221}. These applications highlight the practical value of the model, and future research will continue to explore these concepts.
%
%
\bibliographystyle{splncs04unsort}

\end{document}